\documentclass[twocolumn]{revtex4-1}
\usepackage{graphics,epsfig,amsfonts,amssymb,amsmath,color}\usepackage[normalem]{ulem}
\usepackage{bbm}
\usepackage[mathcal]{euscript}
\usepackage{color}

\begin{document}

\title{Spin qubit manipulation of acceptor bound states in group IV quantum wells}
\author{J.C. Abadillo-Uriel}
\email{jcgau@icmm.csic.es}
\affiliation{Instituto de Ciencia de Materiales de Madrid, ICMM-CSIC, Cantoblanco, E-28049 Madrid (Spain)}
\author{M.J. Calder\'on}
\affiliation{Instituto de Ciencia de Materiales de Madrid, ICMM-CSIC, Cantoblanco, E-28049 Madrid (Spain)}
\date{\today}
\begin{abstract}
The large spin-orbit coupling in the valence band of group IV semiconductors provides an electric field knob for spin-qubit manipulation. This fact can be exploited with acceptor based qubits. Spin manipulation of holes bound to acceptors in engineered SiGe quantum wells depends very strongly on the electric field applied and on the heterostructure parameters. The g-factor is enhanced by the Ge content and can be tuned by shifting the hole wave-function between the heterostructure constituent layers. The lack of inversion symmetry induced both by the quantum well and the electric fields together with the g-factor tunability allows the possibility of different qubit manipulation methods such as electron spin resonance, electric dipole spin resonance and g-tensor modulation resonance. Rabi frequencies up to hundreds of MHz can be achieved with heavy-hole qubits, and of the order of GHz with light-hole qubits. 

\end{abstract}
\maketitle

\section{Introduction}
Solid state spin qubits such as electron spin in quantum dots~\citep{LossPRA1998}, nuclear spin of a donor~\citep{KaneNature1998} or singlet-triplet states of two electron spins~\citep{PettaScience2005} are promising candidates for the creation of a quantum computer with desirable properties like scalability or long coherence times. Group IV semiconductors are expected to have long coherence times; the isotopic purification allows extraordinarily long coherence times in both Silicon and Germanium, making them two of the most promising hosts for spin-qubits~\citep{TyryshkinNatMat2012, MorelloNatureNano2016, SigilitoPRB2016}. In recent years, coherent manipulation of single electron spins has been achieved in both quantum dots and donors~\citep{MorelloNat2010, MauneNature2012, PlaNature2012, VandersypenNatureNano2014}. The nuclear spin of a $^{31}$P donor can also be controlled with very high fidelity, allowing the creation of a two qubit logic gate in silicon~\citep{DzurakNature2015}. Most spin-qubits rely on the use of time dependent magnetic fields to perform operations, but experimentally it is quite difficult to localize a time dependent magnetic field on a single qubit. This makes desirable to look for ways to manipulate the qubit states only by electric fields, like  electric dipole spin resonance (EDSR)~\citep{NowackScience2007, KouwenhovenNature2010} or g-tensor modulation resonance (g-TMR)~\citep{KatoScience2003}.

The search for electric field manipulable qubits has focused the attention in the recent years to high spin orbit systems \citep{NowackScience2007, SalfiPRL2016, SalfiNano2016, KouwenhovenNature2010, PalyiPRL2012, BulaevPRL2007, SzumniakPRL2012}. These systems mix the spin with the orbital degrees of freedom. As the orbital wavefunction is sensitive to electric fields, this mixing allows the possibility of manipulating spins entirely by electric means. In silicon and germanium, the conduction band has a small spin orbit interaction, but in the valence band this interaction can be much stronger. As holes in group IV semiconductors have an orbital momentum $I=1$ and spin $1/2$ in their atomic wavefunctions, in Si and Ge holes can be described by a total effective spin $J=3/2$~\citep{KohnPR1955, PajotSpringer2010}. The p-type orbital momentum is also expected to reduce the hyperfine interaction with the nuclei, reducing the nuclear-spin dephasing of these systems \citep{HigginbothamNanoLett2014}. Due to the spin orbit interaction, the spin can be coupled to phonons~\citep{RuskovPRB2013} and manipulated not only by magnetic but also by electric fields~\citep{SalfiNano2016, SalfiPRL2016}. The quantum confinement has an important role in the spin-orbit interaction in silicon, as it can have an important influence on the ground state mixing between light holes $m_J=\pm 1/2$ and heavy holes $m_J=\pm 3/2$~\citep{KatsarosNatureNano2010}. The manipulation by electric means of holes in silicon has been recently achieved in nanowires~\citep{SilvanoNatureComms2016}. Acceptors in silicon have a small intrinsic T$_d$ symmetry term that allows an extra heavy hole - light hole mixing under the application of electric fields~\citep{BirJPCS1963, BirJPCS1963II}. This T$_d$ symmetry term, together with the lack of inversion symmetry that can be provided through electric fields or by the nanostructure itself, can lead to the enhancement of the spin orbit interaction via a Rashba type interaction. This interaction can create a sweet spot for specific values of the electric field in an EDSR manipulated light-hole acceptor qubit, allowing both fast operations and high coherence times~\citep{SalfiNano2016, SalfiPRL2016}.

A strong spin orbit interaction, together with quantum confinement, can also result in electrically tunable g-factors~\citep{KatsarosNatureNano2010, AresPRL2013}. It has also been shown that the confinement can generate an anisotropy in the g-factor for different states~\citep{ZinovievaPRB2003, BauerAPL2004, LossNanoLett2016}. A heavy hole state in a quantum well has a suppressed g-factor in the plane parallel to the quantum well, while in the case of a light hole the g-factor is suppressed in the perpendicular direction. Both the anisotropy and the tunability of the g-factor are requirements for an electrically manipulated spin qubit by g-TMR~\citep{KatoScience2003, AresAPL2013}. Indeed, the control of g-factors of holes in silicon nanowires has been proven~\citep{VoisinNanoLett2016}.

Good coherence and relaxation times together with the different possibilities of manipulation with electric fields is what make hole systems in silicon an interesting platform for quantum computing. The intrinsic T$_d$ symmetry of the acceptor together with the symmetry reduction due to the quantum confinement allows the possibility of manipulation by electric means. In this work we are interested in exploring acceptors in SiGe heterostructure quantum wells. The different Germanium content of the barrier and quantum well allows g-factor manipulation and will also change the sensitivity to Rashba interacting terms due to a larger Bohr radius. Depending on the type of strain in the Si$_{1-x}$Ge$_x$ quantum well, heavy hole (HH) or light hole (LH) qubits can be defined. For HH qubits, we will explore three different ways of manipulating the acceptors: electron spin resonance with magnetic fields (ESR), electric dipole spin resonance (EDSR) and g-tensor modulation resonance (g-TMR).  For LH qubits, we will focus on pure electric manipulation. To explore all these alternatives, we use an effective mass aproach with the Kohn-Luttinger Hamiltonian for a bulk acceptor in a group IV semiconductor host~\citep{KohnPR1955, BaldereschiPRB1971, LipariSSC1978, LipariSSC1980}. We include the effects of the quantum well barriers, the Bir-Pikus Hamiltonian~\citep{BirJPCS1963, BirJPCS1963II} for including the strain in the quantum well, and the effect of electric and magnetic fields. We will see how strain, which induces a HH-LH splitting, hinders the manipulation by both electric and magnetic fields, while the asymmetry due to the acceptor position within the well or the presence of a vertical electric field facilitates the electric field manipulability. The g-factors dependence on the heterostructure composition is also an important factor for both magnetic and electric field manipulation.

\begin{figure}[h]
\includegraphics[clip,width=0.5\textwidth]{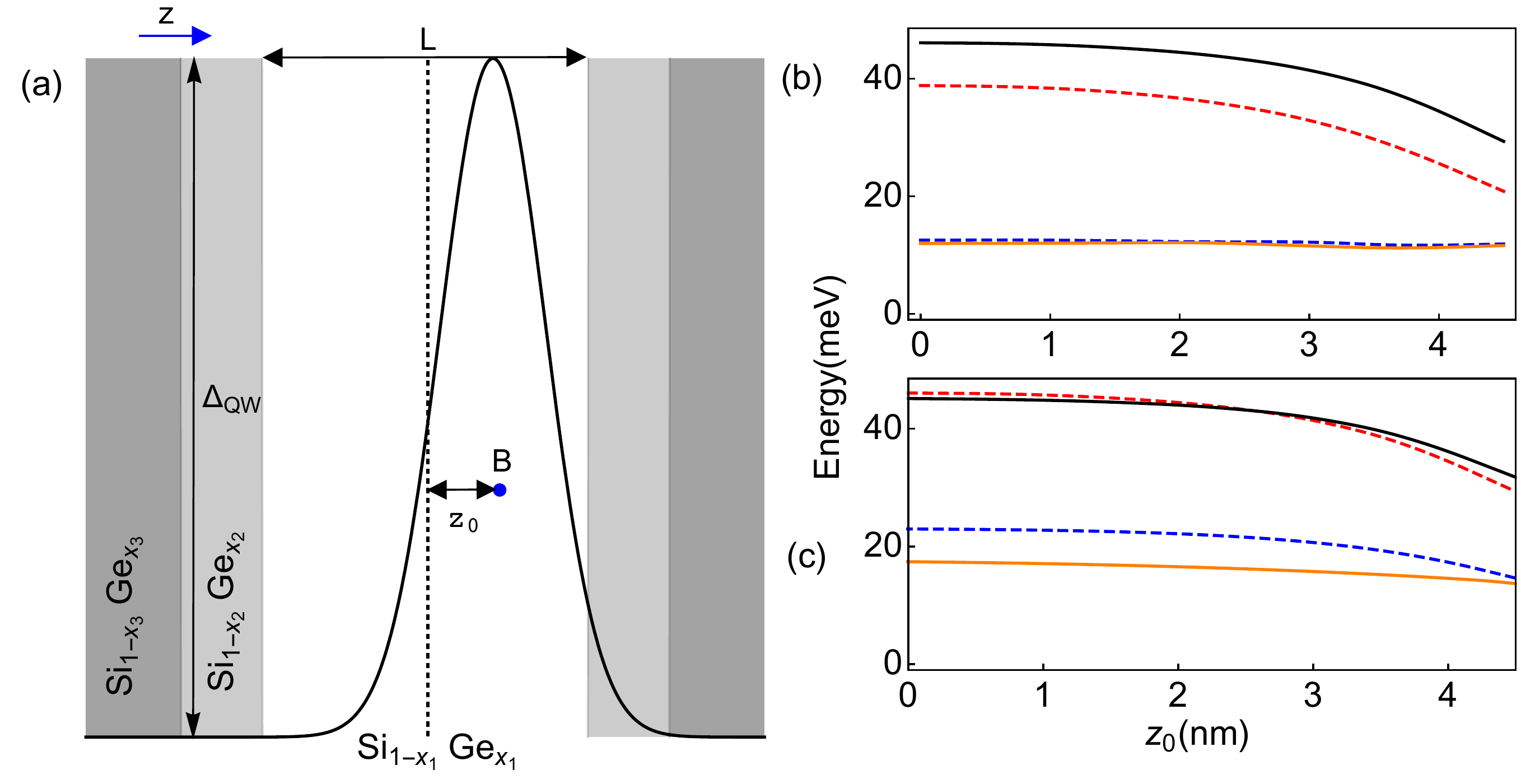}
\caption{\label{fig:scheme} (a) Sketch of the quantum well heterostructure and the bound hole envelope wave-function along the perpendicular $z$ direction. $L$ is the width of the well, $z_0$ determines the acceptor position from the well center, $\Delta_{\rm QW}$ is the barrier height, and $x_1$, $x_2$, and $x_3$ indicate the proportion of Germanium on each of the layers. $x_1>x_2$ for defining the quantum well for holes. $x_3$ determines the sign of the strain on the quantum well and is hence related to whether the doublet ground state is heavy-hole ($x_3=x_2$) or light-hole ($x_3>x_1$) like. (Right) Acceptor energy levels versus $z_0$ for $x_1=0.05$ and $L=10$~nm. No external fields applied in the two cases. Solid (dashed) curves correspond to heavy-hole (light-hole) like states. Note that the energy is calculated from the top of the valence band being the bound states positive in energy. In (b) the ground state is heavy-hole like. The plot corresponds to $x_2=x_3=0$. In (c) the ground state is light-hole like for centred acceptors (small $z_0$) but the level crosses the heavy hole excited one when the acceptor gets close to the barrier due to the effect of quantum confinement. This plot uses $x_2=0$ and $x_3=0.06$. Larger values of $x_3$ would increase the LH-HH splitting. }
\end{figure}

\section{Model}
\label{sec:model}

\subsection{Hamiltonian}
\label{subsec:Hamiltonian}

The acceptor is placed in a quantum well of width $L$, see Fig.~\ref{fig:scheme}.
The Hamiltonian is given by the sum of different contributions
\begin{equation}
\label{TotalHamiltonian}
H=H_{\rm KL}+H_{\rm c}+H_{\rm QW}+H_{\rm BP}+H_{\rm ion}+H_{\rm E}+H_{\rm B}+H_{\rm ic} 
\end{equation}
where $H_{\rm KL}$ is the $6 \times 6$ Kohn Luttinger Hamiltonian~\cite{KohnPR1955} which describes the valence (heavy hole, light hole and split-off) bands in bulk semiconductors. For this term we use the notation and parameters described in Ref.~\cite{AbadilloNanotech2016}.

The acceptor Coulomb potential is given by $H_{\rm c}=e^2/(4\pi\varepsilon_r(z) \varepsilon_0 r)$, with the relative permittivity $\varepsilon_r(z)$ a function of the position in the heterostructure as it is a material dependent parameter. The quantum well potential is described by $H_{\rm QW}=\Delta_{\rm QW}(\Theta (z-L/2)+\Theta (-z-L/2))$, with $\Delta_{\rm QW}$ the energy barrier which depends on the heterostructure composition. The Bir-Pikus Hamiltonian $H_{\rm BK}$\citep{BirJPCS1963, BirJPCS1963II} includes the effect of the strain in the quantum well
\begin{eqnarray}
H_{\rm BP}&=&a\epsilon\mathbbm{1}\nonumber \\
&+&b\left((J_x^2-\frac{5}{4}\mathbbm{1})\epsilon_{xx}+(J_y^2-\frac{5}{4}\mathbbm{1})\epsilon_{yy} 
+ (J_z^2-\frac{5}{4}\mathbbm{1})\epsilon_{zz}\right) \nonumber \\
&+&d/\sqrt{3}\left(\{J_x,J_y\}\epsilon_{xy}
+\{J_y,J_z\}\epsilon_{yz}+\{J_x,J_z\}\epsilon_{xz}\right).
\label{HBP}
\end{eqnarray}
The parameters $a,b$ and $d$ are the deformation potentials, $\bf J$ is the angular momentum, $\epsilon_{ii}$ are the deformation tensor components, and $\mathbbm{1}$ is the identity matrix. The $\epsilon_{ii}$ depend on the relative values of $x_1$, $x_2$ and $x_3$ (see Fig.~\ref{fig:scheme}), the Ge content of each of the layers of the heterostructure.

$H_{\rm ion}$ is the interaction of the acceptor ion with the electric field $\bf E$
\begin{equation}
H_{\rm ion}=p/\sqrt{3}\left(\{J_y,J_z\}{\rm E_x}+\{J_x,J_z\}{\rm E_y}+\{J_x,J_y\}{\rm E_z}\right) \,.
\label{Hion}
\end{equation}
This linear coupling is only possible because the local T$_d$ symmetry of the acceptor central cell does not fulfill the inversion symmetry. This coupling is hence stronger the larger the probability density at the acceptor. It allows for the mixing of HH and LH in the presence of an electric field, independently of the type of confinement. 
The parameter $p$ is an effective dipole moment that can be estimated~\cite{KopfPRL1992} by $p=e\int_0^a F^*(r)r F(r)$ with $a$ the lattice constant of the host material, and $F(r)$ the radial envelope function. In silicon $p=0.26$ Debye.

The electric field interaction with the hole is given by the Stark Hamiltonian $H_{\rm E}=e\mathbf{E}\cdot\mathbf{r}$ whose in-plane components, when the inversion symmetry is lost in the heterostructure, act like a Rashba-type interaction~\citep{SalfiPRL2016, SalfiNano2016}.

The interaction with magnetic fields is given by
\begin{equation}
H_{\rm B}=\mu_{\rm B} \left(g_1(z)\mathbf{B}\cdot\mathbf{J}+g_2(z)\mathbf{B}\cdot\mathcal{J}\right)
\label{MagneticH}
\end{equation}
where $g_1(z)$ and $g_2(z)$ are the linear and cubic bulk g-factor of each material respectively. The operator $\mathcal{J}$ is $\mathcal{J}=(J_x^3,J_y^3,J_z^3)$. The cubic term can be neglected for most of the cases as in general $g_1\gg g_2$~\citep{KopfPRL1992}. However, as the g-factors of confined heavy holes are known to be suppressed for in plane magnetic fields~\citep{BauerAPL2004}, there can be situations in which the cubic g-factors dominate. This cubic term is also important when considering a heavy-hole qubit as it can mix the ground state heavy-hole Kramer doublet through in plane magnetic fields.

Finally, the image charges and hole self-energy are included in $H_{\rm ic}$~\citep{ThoaiPRB1990}. However, this contribution is negligible as the dielectric constants of the well and barrier materials are similar.

To obtain the energies of the bound states in the quantum well we first solve the eigenenergies of the Hamiltonian without the impurity terms, obtaining the top of the valence band in the quantum well. Then the solution of the total acceptor Hamiltonian is calculated from the top of the valence band.

\subsection{Si$_{1-x}$Ge$_x$ Parameters}
\label{subsec:SiGe Parameters}
The valence band description of the Kohn-Luttinger Hamiltonian is parameterized in terms of the Luttinger parameters $\gamma_1$, $\gamma_2$, $\gamma_3$ and the spin-orbit coupling $\Delta_{\rm SO}$, which are tabulated for both silicon and germanium~\citep{BaldereschiPRB1974}. To obtain the values of these parameters for Si$_{1-x}$Ge$_x$ we use the interpolating functions given in Ref.~\citep{Ushakov}. The relaxed lattice constant of a generic Si$_{1-x}$Ge$_{x}$ alloy can be obtained using the interpolating function~\citep{YangSST2004} $a_0(x)=0.541(1-x)+0.5658x-0.00188x(1-x)$, where the lattice parameter increases with the Ge content. The barrier height of the quantum well $\Delta_{\rm QW}$ is related to the valence band offset of the SiGe heterostructure, and is given in Ref.~\citep{YangSST2004}.

The lattice mismatch in the quantum well causes uniaxial strain. For a Si$_{1-x_1}$Ge$_{x_1}$ structure with lattice constant $a_0(x_1)$ grown on a Si$_{1-x_3}$Ge$_{x_3}$ substrate with lattice constant $a_0(x_3)$ the deformation tensor is diagonal with components:
\begin{eqnarray}
\epsilon_{xx}&=&\epsilon_{yy}=\frac{a_0(x_3)-a_0(x_1)}{a_0(x_1)} \nonumber \\
\epsilon_{zz}&=&-2\frac{C_{12}}{C_{11}}\epsilon_{xx}
\end{eqnarray} 
where $C_{12}$ and $C_{11}$ are the elasticity moduli. The uniaxial strain breaks the fourfold degeneracy of the valence band at the $\Gamma$ point  splitting the HH and the LH Kramers doublets. 
For $\epsilon_{xx}>0$ (tensile strain) the top of the valence band has LH character, while it is HH like for $\epsilon_{xx}<0$ (compressive strain). 

\begin{figure}
\includegraphics[clip,width=0.3\textwidth]{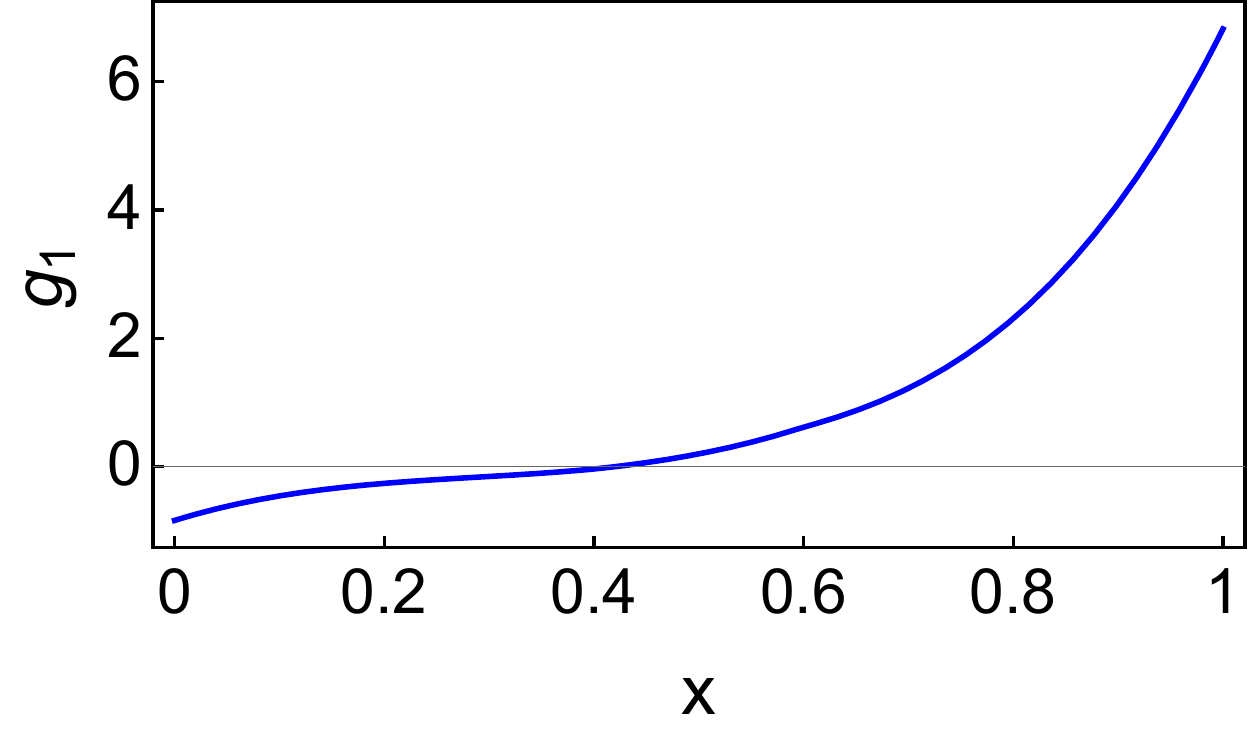}
\caption{\label{fig:gfactorx} Value of the linear g-factor $g_1$ in Si$_{1-x}$Ge$_x$ as a function of the Ge content. The bulk g-factors in Silicon and Germanium have opposite signs, hence around $x=0.4$ the Si$_{1-x}$Ge$_x$ g-factor goes through zero.}
\end{figure}

The deformation potentials $a, b$ and $d$, see Eq.~\ref{HBP}, dielectric constants and elasticity moduli are calculated by linear interpolation~\citep{zhaoJAP1999}. The g-factors are calculated by interpolation of the bulk data from Ref.~\citep{FrajJAP2007}, see Fig.~\ref{fig:gfactorx}.

\subsection{Variational method}
We split the total Hamiltonian Eq.~(\ref{TotalHamiltonian}) into a static Hamiltonian $H_0$ and an interacting Hamiltonian $H_{\rm int}$ 
\begin{eqnarray}
H&=&H_0+H_{\rm int}\nonumber \\ 
H_0&=&H_{\rm KL}+H_{\rm c}+H_{\rm QW}+H_{\rm BP}+H(\rm E_z)_{\rm ion}\nonumber \\ &+&H_{\rm E_z}+H_{\rm B_{static}}+H_{\rm ic} \nonumber \\ 
H_{int}&=&H_{\rm E_{\parallel}}+H_{\rm B_{int}}
\label{interaction}
\end{eqnarray}
$H_0$ includes the contribution of the terms related to the well parameters -- length, uniaxial strain and acceptor position. It also includes static contributions of both the vertical electric field, which due to the acceptor ion term gives an extra mixing of HH and LH, and the static magnetic field, which breaks time reversal symmetry lifting the remaining degeneracy. This static magnetic field is in the perpendicular direction for the HH case, and in the in-plane direction for the LH case. The interacting Hamiltonian $H_{\rm int}$ includes the oscillating in-plane electric and, in the HH case, also in-plane magnetic fields. These terms mix the HH and LH subspaces such that after solving the static Hamiltonian their contribution is mostly off-diagonal in the qubit subspace.

The variational basis set used for solving the static Hamiltonian for the acceptor bound states is
\begin{equation}
|\psi_i(\rho,z,\varphi)\rangle = N_i \rho^{n}e^{-\beta_i\rho}\phi_i(z)e^{iL_z\varphi}|J,J_z\rangle \, ,
\end{equation}
where $N_i$ is the normalization coefficient.
The functions $\phi_i(z)$ are odd and even solutions to the finite quantum well problem, including excited states with different depths outside the quantum well. The set of $\beta_i$ parameters is chosen to be $\beta_1=4$, $\beta_2=2$, $\beta_3=1$ and $\beta_4=0.5$. The number of $\beta_i$ parameters and their value are not as important for the energy convergence as the value of $n$ at which the basis set is truncated. We take $n_{\rm max}=11$. The difference in energy between the calculated ground state using $n_{\rm max}=11$ and $n_{\rm max}=10$ is smaller than $0.1$~meV.

An acceptor bound ground state in bulk is fourfold degenerate (due to the degeneracy of the top of the valence band at the $\Gamma$ point). In a heterostructure this degeneracy is broken both by strain (tensile strain gives rise to a LH ground state while compressive strain leads to a HH ground state~\citep{ThompsonIEEE2006, SunJAP2007}) and quantum confinement (which always favours a HH ground state). A LH ground state may then be produced if the tensile strain splitting overcomes the one produced by the quantum confinement.

\subsection{Schrieffer-Wolff transformation}
$H_0$ is solved variationally, obtaining both the energies and eigenfunctions of the first eight states. Note that, unlike the unstrained bulk acceptor states, the excited states beyond the first two doublets are close in energy (see Fig.~\ref{fig:scheme}) and hence cannot be neglected. The interaction with in-plane electric and magnetic fields is evaluated taking into account this first eight states manifold through a Schrieffer-Wolff transformation up to third order. Due to the lack of inversion symmetry in the well (except when $z_0=0$ and $\rm E_z=0$) and the extra mixing of HH-LH states via the T$_d$ symmetry interaction with electric fields, these in-plane electric and magnetic fields will give off-diagonal terms in the heavy-hole ground state manifold, allowing the manipulation of the qubit state.

\begin{widetext}
We show here the effective Hamiltonians, separating the perpendicular and parallel terms, of only the first four bound states (HH$_1$ and LH$_1$) in the basis $\{3/2,-3/2,1/2,-1/2\}$

\begin{equation}
H_{\perp}^{\rm eff}(\rm E_z, B_z)=\begin{pmatrix}
E_{\rm HH_1}(\rm E_z) & 0 & 0 & -ip\rm E_z \\
0 & E_{\rm HH_1}(\rm E_z) & ip\rm E_z & 0 \\
0 & ip\rm E_z & E_{\rm LH_1}(\rm E_z)  & 0 \\
-ip\rm E_z & 0 & 0 & E_{\rm LH_1}(\rm E_z) 
\end{pmatrix} +\mu_B \rm B_z\begin{pmatrix}
g_{\perp}^{\rm HH_1} & 0 & 0 & 0 \\
0 & -g_{\perp}^{\rm HH_1} & 0 & 0 \\
0 & 0 & g_{\perp}^{\rm LH_1}  & 0 \\
0 & 0 & 0 & -g_{\perp}^{\rm LH_1} 
\end{pmatrix} \, .
\label{H0}
\end{equation}
Where the values $g_{\perp}^{\rm HH_1}$, $g_{\perp}^{\rm LH_1}$ are the perpendicular g-factors of the first HH and LH bound states respectively. These g-factors include the contributions of both the linear and cubic g-factors. The energies $E_{\rm HH_1}$ and $E_{\rm LH_1}$ depend on $L$, $z_0$ and the electric field applied in the $z$ direction E$_z$.
Note that for the numeric results we are considering an $8\times 8$ effective Hamiltonian. The last four states are important for the quantitative results as they can be close to both the first LH$_1$ and HH$_1$ states, however the qualitative picture can already be understood in terms of an effective $4\times 4$ Hamiltonian involving the Kramer doublets of the first HH$_1$ and LH$_1$ states.

In the same basis, the effective Hamiltonian of the in-plane terms is, showing only the linear terms for simplicity,

\begin{equation}
H_{\parallel}^{\rm eff}(\mathbf{E}_\parallel, \mathbf{B}_{\parallel})=\begin{pmatrix}
0 & 0 & -ip\rm E_{+} +\alpha \rm E_{-} & 0 \\
0 & 0 & 0 & -ip\rm E_{-} -\alpha \rm E_{+} \\
ipE_{-} +\alpha \rm E_{+} & 0 & 0 & 0 \\
0 & ipE_{+} -\alpha \rm E_{-} & 0 & 0 
\end{pmatrix}
+\mu_B \rm \begin{pmatrix}
0 & \tilde g'_{\parallel} B_{+}& \frac{\sqrt{3}}{2} \tilde g_{\parallel} B_{-}& 0 \\
\tilde g'_{\parallel} B_{-}& 0 & 0 & \frac{\sqrt{3}}{2}\tilde g_{\parallel} B_{+}\\
\frac{\sqrt{3}}{2}\tilde g_{\parallel} B_{+} & 0 & 0 & 
\tilde g_{\parallel} B_{-}\\
0 & \frac{\sqrt{3}}{2}\tilde g_{\parallel} B_{-} & \tilde g_{\parallel} B_{+} & 0 
\end{pmatrix} .
\label{Hint}
\end{equation}
Here $\alpha$ is the Rashba coupling parameter, and the g-factors $\tilde g_{\parallel}$ and $\tilde g'_{\parallel}$ are the effective linear and cubic g-factors, respectively.
\end{widetext}

For a HH qubit we will consider the Hamiltonian Eq.~(\ref{H0}) as an effective static Hamiltonian, with Eq.~(\ref{Hint}) the interacting Hamiltonian. For a LH qubit however, the static magnetic field will be in the in-plane direction to ensure the existence of sweet spots~\citep{SalfiPRL2016}, and it will be manipulated only by electric means.

From Eq.~(\ref{Hint}) it can be seen that the linear g-factor does not couple heavy-hole states. This occurs because it is only coupled to the linear spin operator $\mathbf{J}$, see Eq.~\ref{MagneticH}. Under compressive strain in the SiGe quantum well, the HH and LH subbands are separated by a few meV, and the ground state g-factor in the in-plane directions is suppressed as it is off-diagonal, being proportional to the HH-LH mixing.  This implies that the linear in-plane g-factor in the ground state manifold is tunable through the electric field mixing of HH and LH.

On the other hand, the cubic g-factor couples magnetic fields with the operator $\mathcal{J}=(J_x^3,J_y^3,J_z^3)$. This third order spin operator is the only term in the total Hamiltonian Eq.~(\ref{TotalHamiltonian}) that directly couples HH Kramer doublets. Although the cubic g-factor is small, it is not negligible when the linear g-factor is suppressed and can even be the dominant term for an in-plane magnetic field.

Regarding the interaction with electric fields, in the effective static Hamiltonian Eq.~(\ref{H0}) it can be seen that non-zero vertical electric fields already mix the HH and LH states through the off-diagonal T$_d$ symmetry interaction with the acceptor ion. This extra mixing is quite important in the case of electric manipulation of a HH state, as it allows the interaction with in-plane electric fields via both the Rashba and the T$_d$ symmetry terms in Hamiltonian Eq.~(\ref{Hint}).

\section{Results}
\label{sec:results}

By solving Eq.~(\ref{TotalHamiltonian}) we obtain the parameters of the effective Hamiltonian $p$, $\alpha$, $\tilde g_\parallel$, $\tilde g_\perp$. The effective dipole moment $p$ is proportional to the probability density in the central cell region and hence decreases when $x$ increases as the quantum well becomes more of Germanium type and the hole wavefunction in Ge is more widespread than in Si. The other dipole moment $\alpha$ is related to the lack of inversion symmetry and it grows when the hole wavefunction is deformed away from the center of the quantum well.

By choosing the right heterostructure, the ground state can have a heavy hole or light hole character (see Fig.~\ref{fig:scheme}). HH and LH acceptor qubits in Si have been discussed in Refs.~\citep{SalfiPRL2016,SalfiNano2016}. Here we focus on the different opportunities for spin manipulation that the HH or LH acceptor ground states offer in SiGe based quantum wells. 

 The obtained effective g-factors are anisotropic. For HH states, $g_{\parallel}^{\rm HH_1}$ is suppressed while $g_{\perp}^{\rm HH_1}$ remains finite. The opposite is true for LH states. These g-factors change with $x$ as expected from the evolution from the Si to the Ge g-factors (see Fig.~\ref{fig:gfactorx}). The g-factors are also affected by the electric fields via the spin orbit interaction. However, in a heterostructure, the dominant dependence is given by the wave-function density on the different layers, as the g-factors of the quantum well and the barriers are not equal. This wave-function density depends on the acceptor position within the well and it can be tuned by $E_z$.

\subsection{Heavy-hole ground state}
Typically, p-type SiGe quantum wells are compressively strained ($x_2<x_1$). In this case both the quantum confinement and the strain favor the HH states over LH states, see Fig.~\ref{fig:scheme}(b), with a few meV HH-LH splitting. As the HH in-plane g-factors are known to be suppressed under these circumstances we will consider a perpendicular magnetic field to split the HH Kramer doublet. The interacting Hamiltonian is then Eq.~\ref{Hint}, from which we can already tell that the two HH states can only be coupled to first order for in-plane magnetic fields. Manipulating through electric fields then is subject to HH-LH coupling. From Eq.~(\ref{H0}) it can be seen that vertical electric fields can increase this coupling via the local T$_d$ symmetry term, however the interaction will be inversely proportional to the HH-LH energy separation.

\subsubsection{Electron Spin Resonance}
\begin{figure}
\includegraphics[clip,width=0.5\textwidth]{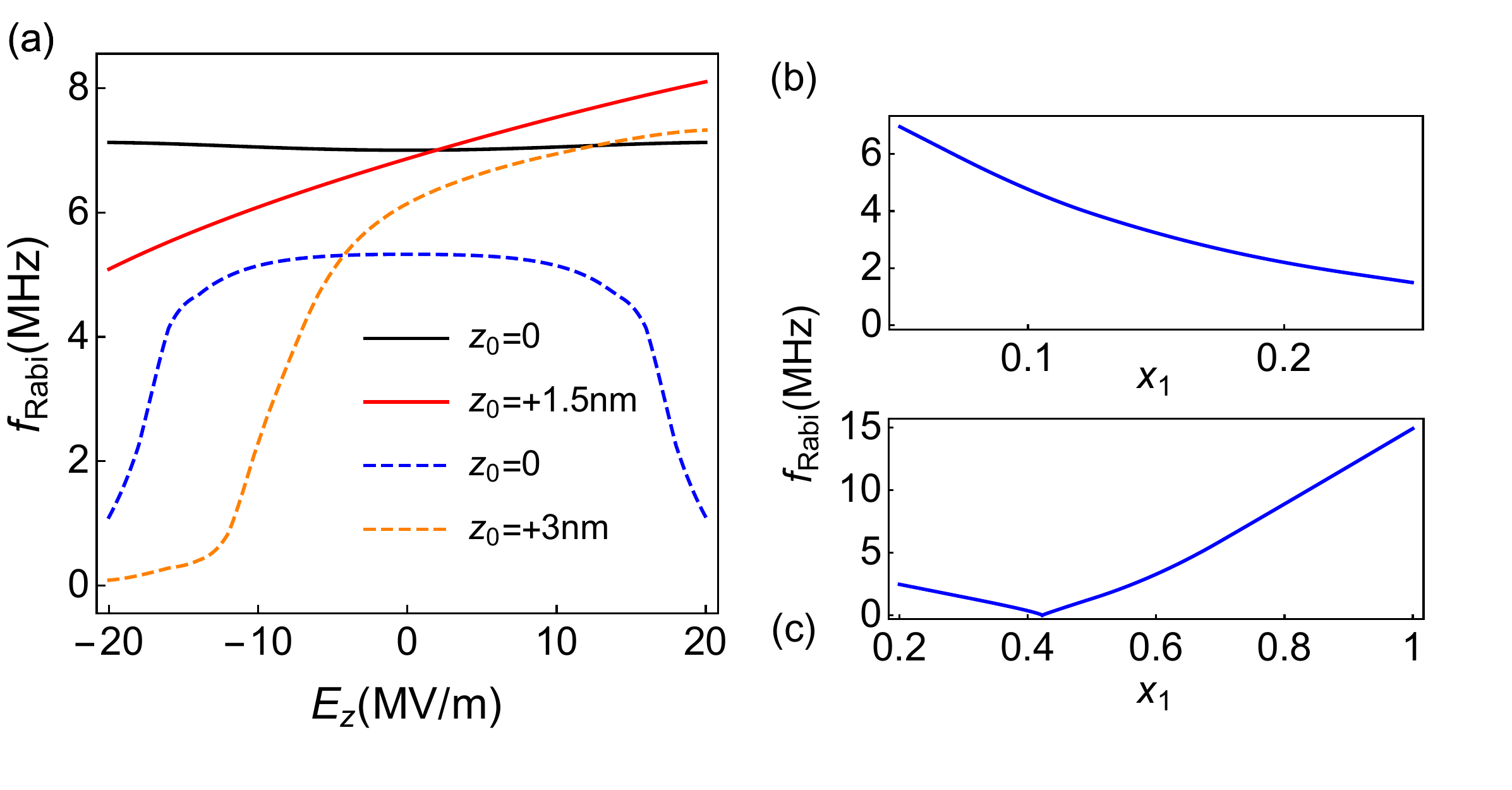}
\caption{\label{fig:ESR} ESR Rabi frequency (a) as a function of the perpendicular electric field for $L=5$ nm (solid lines) and $L=10$ nm (dashed lines) and different values of $z_0$. $x_1=0.05$, and $x_2=x_3=0$; (b) for $L=10$ nm as a function of $x_1$ with $x_2=x_3=0$, $z_0=0.6 \frac{L}{2}$ and $E_z=10$ MV/m; (c) same as (b) with $x_3=x_2=x_1-0.05$. The amplitude of the in-plane oscillating magnetic field applied is $50$ mT and the constant magnetic field applied in the z-direction is $1$ T.}
\end{figure}

Spin manipulation can be readily achieved by applying oscillating in-plane magnetic fields, see Eq.~\ref{Hint}, which directly couple the two heavy hole states proportionally to $\tilde g_\parallel$. Therefore, the Rabi frequency can be enhanced by increasing the in-plane g-factor. As the quantum well and the quantum barrier have different g-factors, increasing the density of the hole wavefunction in the region with a larger g-factor in absolute value will also increase the Rabi frequency of the ESR manipulation. Here we evaluate the effect on the in-plane effective g-factors of three different parameters: The proportion of Germanium $x$ in the SiGe alloy, the applied perpendicular electric field E$_z$, and the position of the acceptor within the quantum well $z_0$.

The dependence of the (bulk) g-factor on the Germanium content $x$ is shown in Fig.~\ref{fig:gfactorx}. This implies that, for instance, for $x_1<0.4$ and $x_2<x_1$, increasing the wavefunction density in the barrier enhances the g-factor. The opposite happens for $x_1>0.6$ and $x_2<x_1$. On the other hand, as $x_1-x_2$ increases so does the barrier height due to the strain, making it harder for the wavefunction to penetrate the barrier. $z_0$ also affects the penetration in the barriers: the closer the acceptor is to the barriers, the larger the density probability in them.

The role of E$_z$ is more complex. On one hand, it can modulate the wave-function probability density in the different layers. However, the electric field has more consequences on the acceptor physics as it changes the HH-LH mixing through the T$_d$ symmetry term (the $ip \rm E_z$ components in Eq.~\ref{H0}). In principle, increasing E$_z$ would increase the linear g-factor but the effective dipole moment $p$, which is proportional to the  wave function probability density near the acceptor, can be simultaneously reduced, limiting the effect of this term in the total Hamiltonian. $p$ can also be affected by $L$ (smaller widths increases probability density) and by $x$ (as the Bohr radius in Ge is larger than in Si).

Fig.~\ref{fig:ESR} illustrates the previous remarks. For small values of $L$ ($L=5$ nm for the solid lines in Fig.~\ref{fig:ESR}(a)) the effect of E$_z$ is small as the wave function is more constrained. Off-centered acceptors ($z_0 \neq 0$) are more easily manipulated by electric fields which push the wave-function towards the farthest away interface. With Si barriers ($x_2=0$), Fig.~\ref{fig:ESR}(b), the Rabi frequency is suppressed as a function of $x_1$ but a significant enhancement can be achieved in all-SiGe heterostructures, as Si$_{1-(x_1-0.05)}$Ge$_{x_1-0.05}$/Si$_{1-x_1}$Ge$_{x_1}$/Si$_{1-(x_1-0.05)}$Ge$_{x_1-0.05}$,  once $x_1>0.6$, as shown in Fig.~\ref{fig:ESR}(c). In summary, the best conditions to enhance the ESR Rabi frequency are achieved by quantum wells with large $x_1$ and by choosing the acceptor position and electric fields such that most of the wavefunction density is in the region with larger g-factor. Wider quantum wells make it easier to tune the Rabi frequency with electric fields.

\subsubsection{Electric Dipole Spin Resonance}

\begin{figure}
\includegraphics[clip,width=0.5\textwidth]{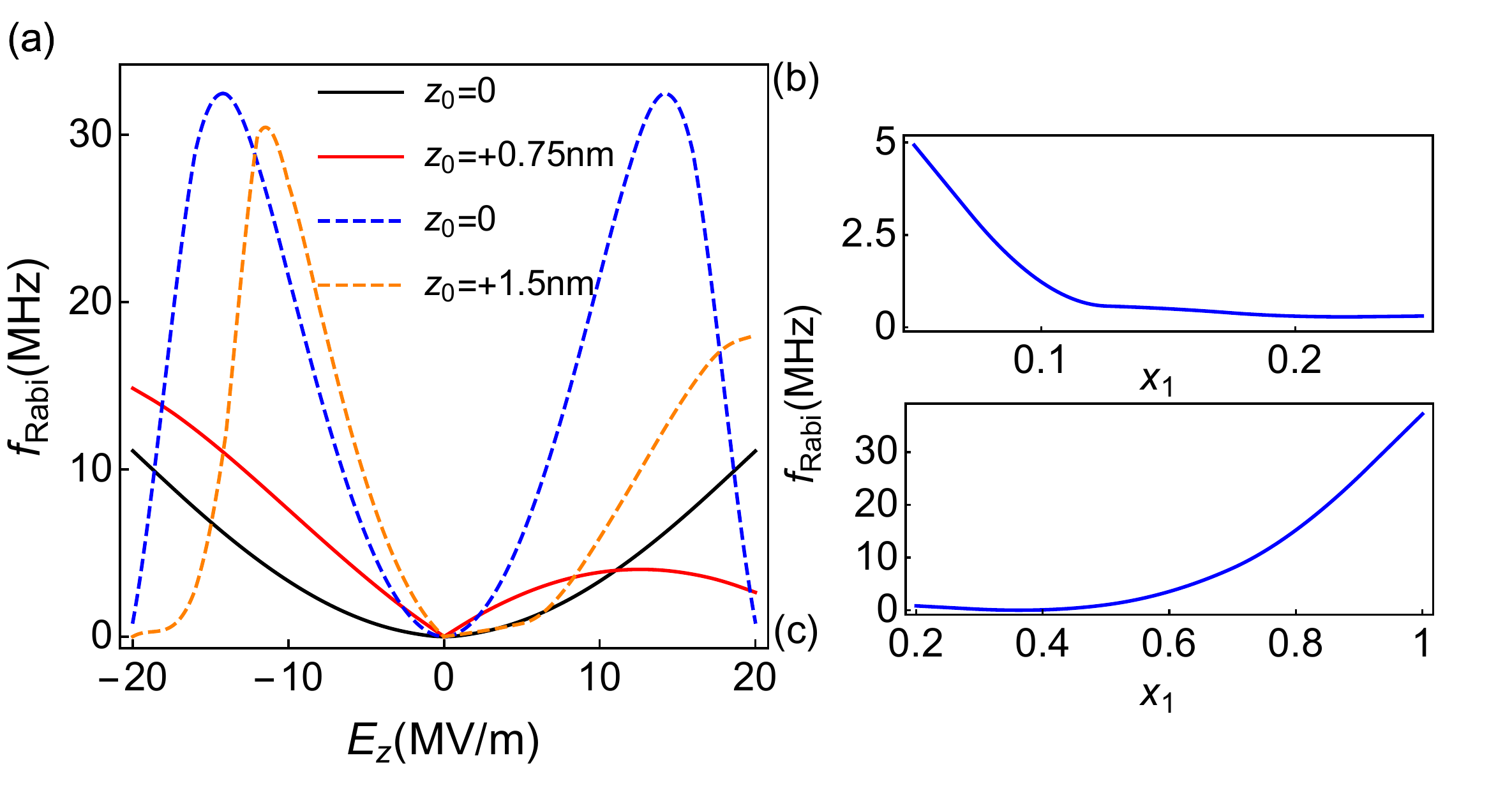}
\caption{\label{fig:EDSR} EDSR Rabi frequency as (a) a function of the perpendicular electric field for $L=5$ nm (solid lines) and $L=10$ nm (dashed lines) and different values of $z_0$. $x_1=0.05$, and $x_2=x_3=0$; (b) for $L=10$ nm as a function of $x_1$ with $x_2=x_3=0$, $z_0=0.6 \frac{L}{2}$ and $E_z=10$ MV/m; (c) same as (b) with $x_3=x_2=x_1-0.05$. The in-plane oscillating electric field is taken as $50$~kV/m and the constant magnetic field applied in the z-direction is $1$ T.}
\end{figure}

Spin-orbit interaction provides us with an electric knob to manipulate spins. The spin-orbit interaction is produced by the inversion symmetry breaking (Rashba $\alpha$ terms in Eq.~\ref{Hint}) in the heterostructure and by the acceptor T$_d$ symmetry ($p$ terms in Eq.~\ref{Hint}).  These terms induce an extra mixing between the HHs and LHs allowing a purely electric field manipulation with in-plane oscillating fields.

The HH-LH coupling is limited by excessive strain and by large $x_1-x_2$, both splitting the HH-LH manifolds by several meV and hence reducing drastically the EDSR term. As shown in Fig.~\ref{fig:EDSR} the absence of a perpendicular electric field $E_z$ leads to zero coupling.

The Rashba term gets stronger by reducing the inversion symmetry. This symmetry is broken by placing the acceptor on an off-centered position and by applying a perpendicular electric field. The barrier height can also increase the Rashba coupling but in exchange this implies higher strain and lower HH-LH coupling. The quantum well width can also be important: the wider the quantum well, the more room the wavefunction has to move, allowing higher Rashba couplings. Another interesting factor is the Bohr radius of the material. Contrary to the T$_d$ term, which is suppressed when the Bohr radius is big (namely, for large Ge content), the Rashba term is enhanced due to the higher sensitivity to the lack of inversion symmetry induced by one of the barriers.

Fig.~\ref{fig:EDSR} (b) and (c) show the dependence of the EDSR Rabi frequency on $x_1$ for a Si and a Si$_{1-(x_1-0.05)}$Ge$_{x_1-0.05}$ barrier, respectively. A large Ge content in the well, together with a small $x_1-x_2$ difference, gives rise to a significant enhancement of the Rabi frequency.

\subsubsection{g-Tensor Modulation Resonance}

\begin{figure}
\includegraphics[clip,width=0.5\textwidth]{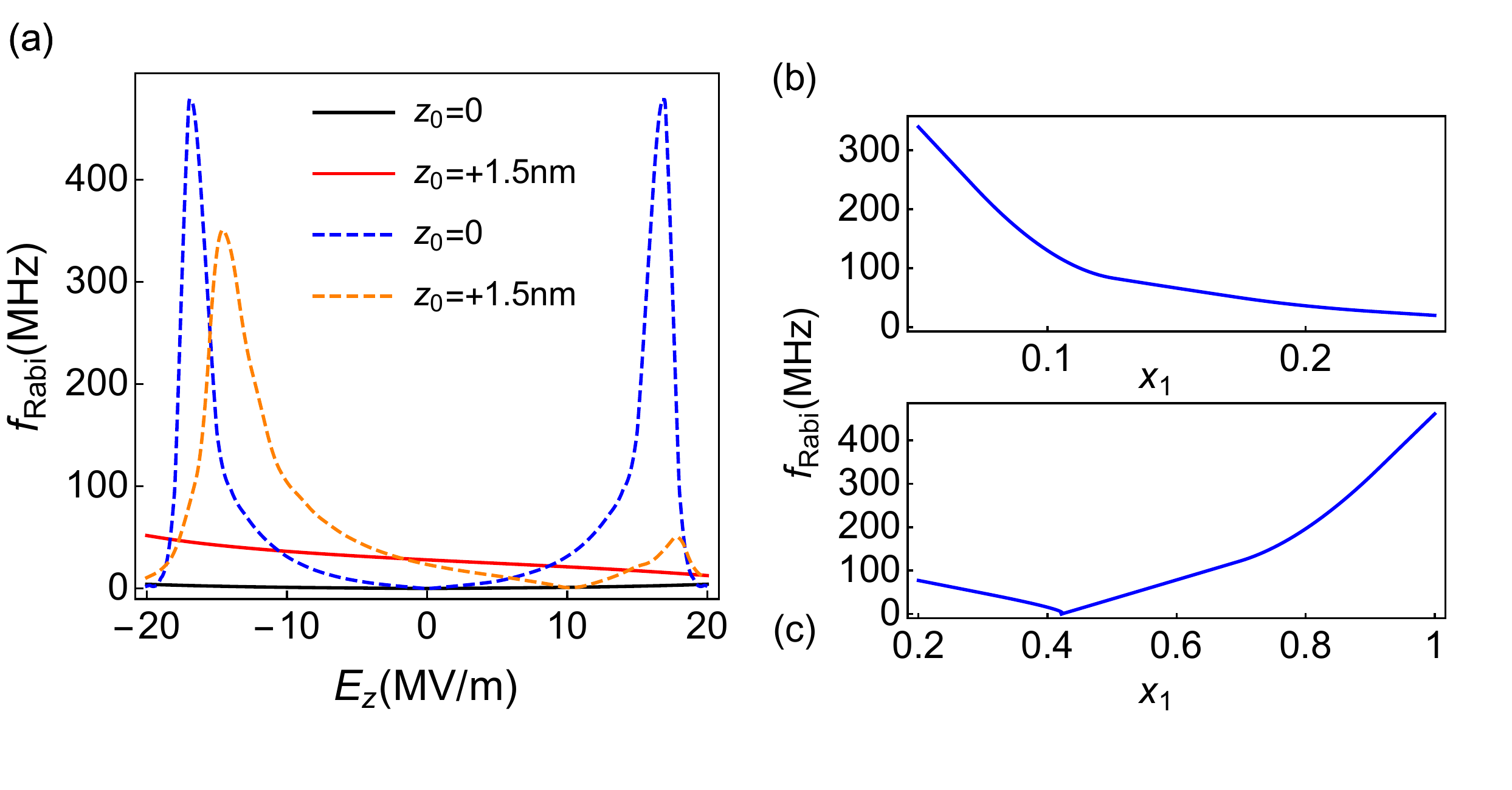}
\caption{\label{fig:GTMR} g-TMR Rabi frequency as (a) a function of the perpendicular electric field for $L=5$ nm (solid lines) and $L=10$ nm (dashed lines) and different values of $z_0$. $x_1=0.05$, and $x_2=x_3=0$; (b) for $L=10$ nm as a function of $x_1$ with $x_2=x_3=0$, $z_0=+3$ nm, $E_z=-10$ MV/m, and $|\mathbf{B}|=1$T; (c) same as (b) with $x_3=x_2=x_1-0.05$.}
\end{figure}

The quantum confinement and the strain suppresses the in-plane g-factor in comparison to the out of plane g-factor of heavy-holes, creating an anisotropy in the g-tensor. This anisotropy depends on applied perpendicular electric fields by means both of the spin-orbit interaction and the different content of Ge on the heterostructure layers.
The dependence of the Rabi frequency on the g-factor modulation g-TMR by E$_z$ is given by~\citep{AresAPL2013} 

\begin{equation}
f_R^{gTMR}=\frac{\mu_B E_{ac}}{2h}\bigg[ \frac{1}{g_{\parallel}}\left( \frac{\partial g_{\parallel}}{\partial E_z}\right)-\frac{1}{g_{\perp}}\left( \frac{\partial g_{\perp}}{\partial E_z} \right) \bigg]\frac{g_{\parallel}g_{\perp}|B|}{|g_{\parallel}|+|g_{\perp}|}
\label{gtmr}
\end{equation}
where $E_{ac}$ is the oscillating component of the applied perpendicular electric field, assumed to be small enough to consider a linear dependence of the g-tensor. This oscillating field is superimposed to the static vertical electric field E$_z$ (already considered in the Hamiltonian $H_0$) and is taken as $E_{ac}=1$~MV/m in the following. The magnetic field is applied in a direction that maximizes the Rabi frequency, $\theta=\arctan\bigg(\sqrt{\frac{g_{\parallel}}{g_{\perp}}}\bigg)$
where $\theta$ is defined with respect to the in-plane direction \citep{AresAPL2013}. As $g_{\parallel}<g_{\perp}$, $\theta\approx 0$.

In order to enhance the g-TMR Rabi frequency, the derivatives of the g-factors with respect to the electric field E$_z$ have to be maximised. This condition is fulfilled for large values of $x_1$ and small $x_1-x_2$ as Ge g-factors are larger. A narrow quantum well would restrain the effect of E$_z$ on the wave-function so wide quantum wells are more desirable. For the same reason, off-centered acceptors give larger frequencies. These results are summarised in Fig.~\ref{fig:GTMR} where it is also patent that the frequencies achieved with this method are at least one order of magnitude larger than with ESR or EDSR.

\subsection{Light-hole ground state}
\begin{figure}
\includegraphics[clip,width=0.3\textwidth]{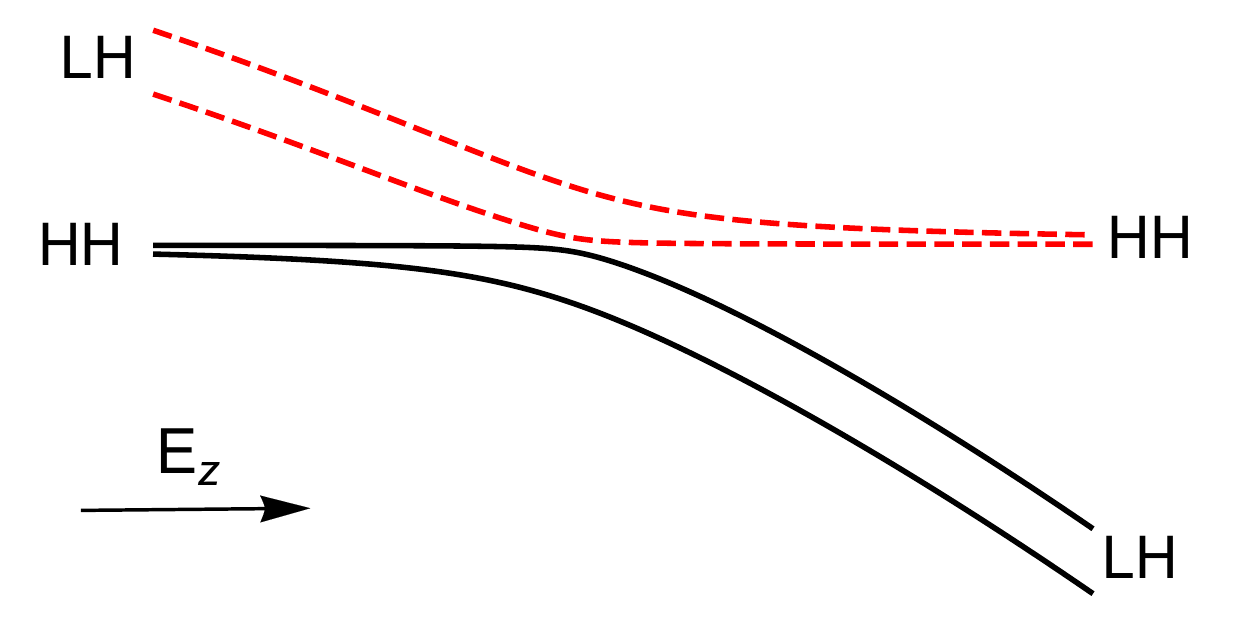}
\caption{\label{fig:LHscheme} Sketch of the first four energy levels under an in-plane constant magnetic field as a function of an electric field $E_z$. The ground state is of LH character for $E_z=0$ due to tensile strain. For larger values of $E_z$, due to the proximity to a barrier interface, the quantum confinement competes with strain and eventually the ground state becomes of HH character. The anticrossing occurs due to the LH-HH mixing induced by the applied fields. For simplicity, the g-factor is assumed to be constant. In practice, it can decrease (increase) as a function of $E_z$ in the concave (convex) case.}
\end{figure}

To form a p-type quantum well with SiGe, $x_2<x_1$ is required. If we take a substrate with different Germanium content $x_3$ (see Fig.~\ref{fig:scheme}) it is possible to get a tensile strained p-type quantum well when $x_3>x_1$. In this case, the splitting caused by the tensile strain competes with the one from quantum confinement such that when the strain is large enough the ground state in the quantum well is of LH nature. Here we will focus on cases where the ground state is LH but the HH-LH splitting can be tuned such that the ground state can become of HH character, see Fig.~\ref{fig:scheme}(c). Proximity to the interface with the barrier (by choosing a particular $z_0$ or by applying large electric fields) tend to favour a HH ground state. This LH-HH proximity can lead to sweet spots in the LH qubit subspace~\citep{SalfiPRL2016} when in-plane magnetic fields are applied to break the Kramers doublet degeneracy. 

As the magnetic field here is applied in the $x-y$ plane and gives rise to an off-diagonal interaction, the LH qubit subspace gets mixed. In the qubit subspace then, neither in-plane nor out-of-plane magnetic fields will give purely off-diagonal interaction. Moreover, as the first four states are close in energy and have different g-factor dependences, the g-TMR manipulation gives rise to complex four state dynamics,  which is beyond the scope of this work. Therefore, for a LH qubit we focus only on purely electric field manipulation through EDSR. 

 The main difference with previous work~\cite{SalfiPRL2016,SalfiNano2016} is the electric field dependence of the g-factor. Depending on the behavior of the g-factor as a function of the electric field (which can push the wave-function inside the barriers) we will distinguish two cases: The convex LH qubit, where the g-factor grows when increasing the density wavefunction within the barriers (for $x_1 \lesssim 0.4$), and the concave LH qubit where the g-factor decreases when increasing the density wavefunction in the barriers (for $x_1 \gtrsim 0.4$). This behavior shows in the Larmor frequency and affects the conditions for the sweet spots.

\subsubsection{Convex LH qubit}
\begin{figure}[h]
\includegraphics[clip,width=0.5\textwidth]{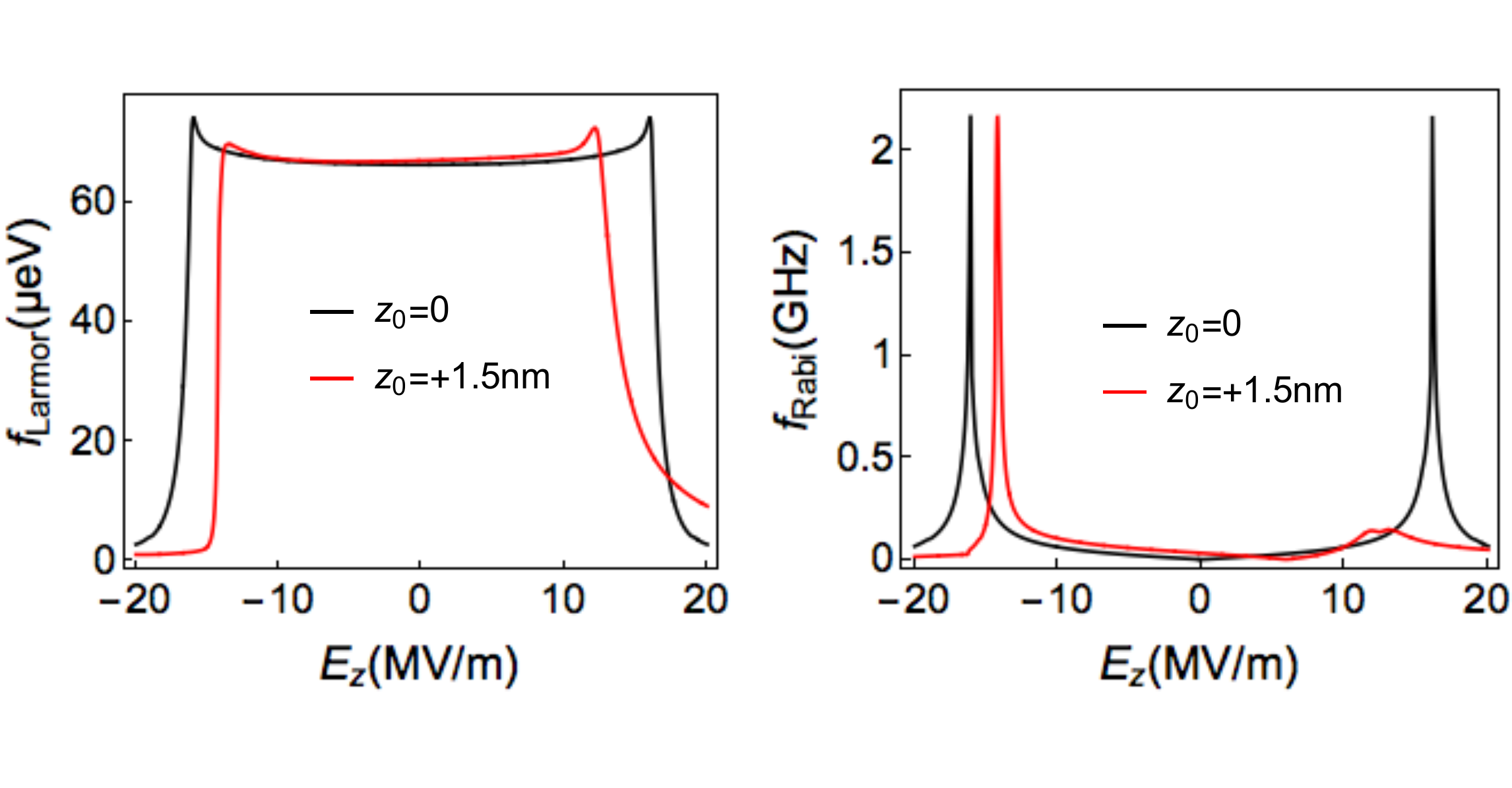}
\caption{\label{fig:LHconvex} (Left) Larmor frequency of the convex LH qubit in $\mu$eV as a function of the electric field for $x_1=0.1$, $x_2=0$, $x_3=0.06$, $B=1$~T, $L=10$~nm. (Right) Rabi frequency of the LH qubit for the same parameters as (Left) and assuming an in-plane oscillating electric field of 50kV/m.}
\end{figure}

The Larmor frequency inherits the convex behavior of the g-factor as long as the ground state is LH (namely, for small values of $E_z$). The implications are shown in Fig.~\ref{fig:LHconvex}: there is a sweet spot at the minimum of the g-factor and from there the qubit frequency grows due to both the g-factor increasing and the interaction with the T$_d$ symmetry term. For larger values of $E_z$, there is a HH-LH anticrossing at which the g-factor decreases because the HH in-plane g-factor is suppressed, giving rise to another sweet spot. This last sweet spot appears both at positive and negative electric fields due to the symmetry of the quantum well. In total there might be up to three sweet spots, with positions that depend on several parameters, particularly the acceptor position $z_0$. 

Regarding the Rabi frequencies, the off-diagonal terms in the qubit subspace come mostly from the Rashba interaction so they grow by reducing the inversion symmetry, becoming maximal at the HH-LH anticrossing. Near the sweet spots, this gives rise to very large Rabi frequencies compared with the HH qubit ones.

\subsubsection{Concave LH qubit}
\begin{figure}[h]
\includegraphics[clip,width=0.5\textwidth]{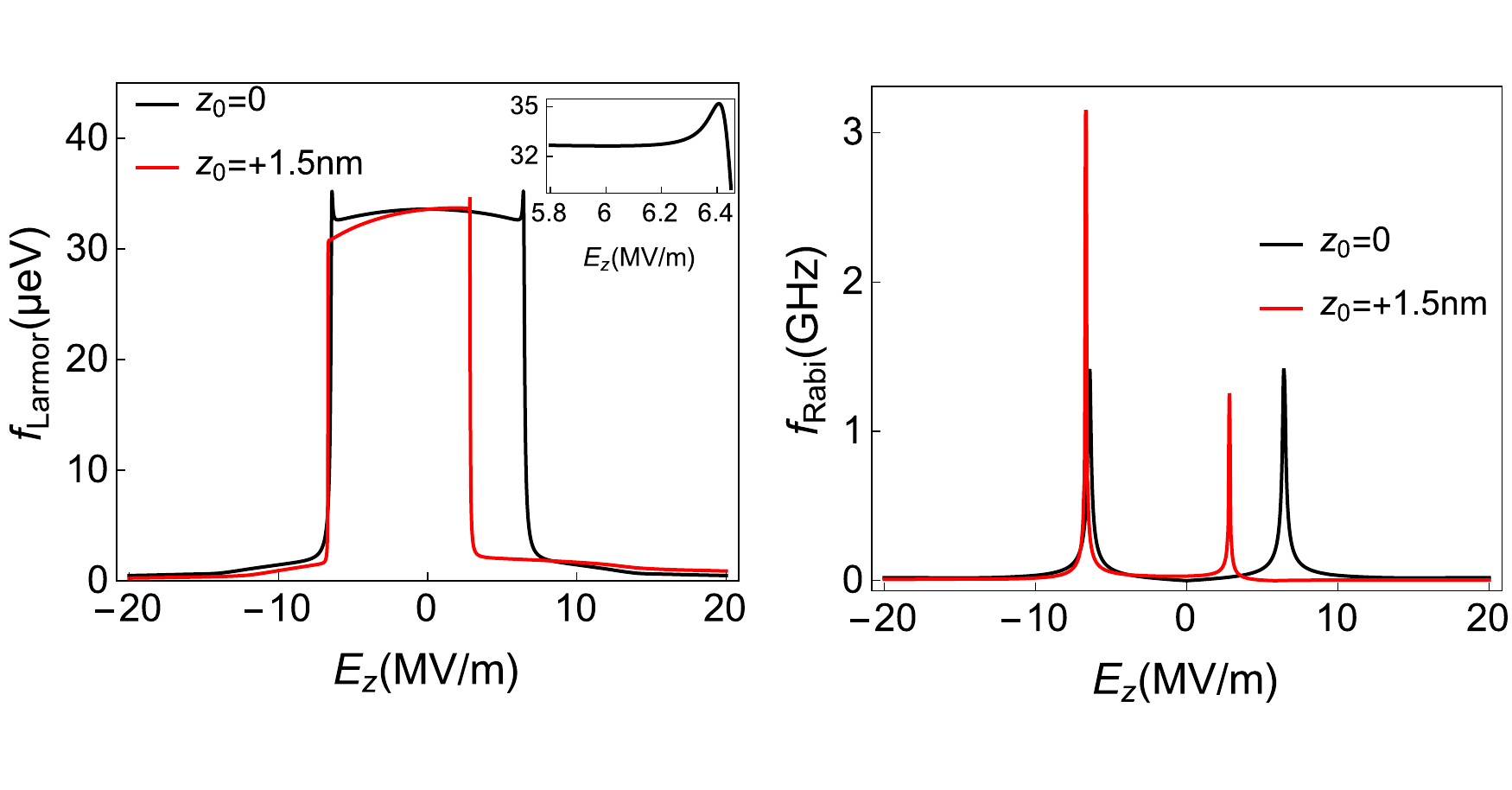}
\caption{\label{fig:LHconcave} (Left) Larmor frequency of the concave LH qubit in $\mu$eV as a function of the electric field for $x_1=0.8$, $x_2=0.7$, $x_3=0.82$, $B=0.1$~T, $L=10$~nm. (Right) Rabi frequency of the LH qubit for the same parameters as (Left).}
\end{figure}

In this case the Larmor frequency inherits the concave behavior of the g-factor. The derivative of the Larmor frequency has now two opposite contributions as a function of the electric field: the decreasing g-factor, and the small positive contribution from the T$_d$ symmetry. If the contribution to the derivative of the g-factor is larger than the contribution of the T$_d$ term, which is the case for high magnetic fields, there is only one sweet spot which corresponds to the maximum of the g-factor. When the contribution from the g-factor is small such that the T$_d$ term can partially surpass it near the anticrossing, there will be two extra sweet spots. In total there can be up to five sweet spots: at the maximum of the g-factor, when the T$_d$ term surpasses  the g-factor in the derivative, and at the final decrease just before the anticrossing due to the mixing with the HH component. Note that the last two might appear both for positive and negative $E_z$ due to the quantum well symmetry. This kind of behavior can be seen in the case $z_0=0$ in Fig.~\ref{fig:LHconcave} where there are five sweet spots. The five sweet spots appear only for small enough magnetic fields: for instance, in the case of Fig.~\ref{fig:LHconcave}, with $x_1=0.8$ and $B=0.1$~T, in the case of negative electric field and $z_0=+1.5$nm, the five sweet spots cannot be attained. Higher $x_1$ would increase the effective g-factor, reducing even further this maximum magnetic field. On the other hand, a smaller $x_1$ could allow higher magnetic fields. 

The Rabi frequencies have the same behavior as in the convex case: the reduction of inversion symmetry clearly increases the Rabi frequencies, being maximal near the sweet spots. The numbers however can be larger as the Rabi frequency of the EDSR in a LH qubit is directly proportional to the Zeeman field, and in this case the effective g-factors are much larger.

\section{Discussion}

For the HH qubit, the Rabi frequencies obtained by electric and magnetic field manipulation are of the order of MHz, see Figs.~\ref{fig:ESR}-\ref{fig:GTMR}. This manipulation frequency has to be benchmarked against relaxation and coherence times in order to get an estimate of the number of qubit rotations allowed before losing coherence. Estimates of the relaxation times using the formula for small temperature phonon-induced spin relaxation of acceptor heavy hole qubits in Ref.~\citep{SalfiNano2016}, and using the relevant parameters for the considered materials \citep{FischettiJAP1996}, are of the order of miliseconds in the worst case scenario (small quantum well barriers) and can be significantly improved by increasing $x_1-x_2$ as the relaxation times are directly proportional to the HH-LH splitting.
Experiments in natural Ge/Si nanowires double quantum dots have measured coherence times in the order of tenths of microseconds~\citep{HigginbothamNanoLett2014}, more than one order of magnitude longer than for III-V semiconductors. These measurements  show a dominant nuclear-spin dephasing, but isotopic purification is expected to make charge noise the dominant source of dephasing.

The Larmor frequency is, for the HH qubit, mostly determined by the Zeeman splitting, other factors such as the HH-LH mixing due to the spin-orbit terms negligible in comparison. This means that the largest contribution to charge noise dephasing comes from the fluctuations in the g-factor. The regions in which the g-factors vary more strongly under the effect of an electric field are also those in which the qubit is more sensitive to charge noise dephasing. On the other hand, as the quantum well is symmetric, there is a value of $E_z$ at which the g-factor is not sensitive to variations of the electric field. This value is $E_z=0$ when the acceptor is placed at the center of the quantum well. Therefore, the best conditions for g-TMR coincide with a large effect of charge noise dephasing and a compromise between the two has to be achieved.

Regarding the LH qubit, there are two desirable properties: strong couplings that allow fast EDSR manipulation together with the existence of sweet spots~\cite{SalfiPRL2016}. The estimated Rabi frequencies are much higher in the LH qubit than in the HH one. In return, relaxation and coherence times are expected to be much smaller. For instance, applying the phonon-induced spin relaxation formula for a light-hole qubit~\cite{SalfiPRL2016}, we get $T_1$ of the order of $\mu$s. Up to five sweet spots can be achieved depending on the heterostructure composition. In the case in which the g-factor is smaller in the barrier (concave LH qubit), the sweet spots near the anticrossing are very close to each other and a better resilience against electric field noise is expected.

\section{Conclusions}
We have calculated the Rabi frequencies of three different ways of manipulating a single qubit for a hole bound to an acceptor inside a quantum well. Depending on the strain conditions, the acceptor ground state has a heavy-hole or light-hole character. The results show that it is possible to get Rabi frequencies for a HH ground state in the range of MHz for the three different manipulation methods while it is possible to reach the GHz in the case of an electrically manipulated LH state. 

 The Rabi frequency of the electron spin resonance of a HH state can be enhanced by increasing the g-factor in the heterostructure (this can be achieved by a large Ge content) and by raising the hole density wave-function in the barriers, which have a larger g-factor. It is hence interesting to use small barriers, for instance 
Si$_{1-x_2}$Ge$_{x_2}$/Si$_{1-x_1}$Ge$_{x_1}$/Si$_{1-x_2}$Ge$_{x_2}$ quantum wells with $x_2=x_1-0.05$. An electric field can shift the hole wave-function within the heterostructure and hence can be applied to increase or decrease the Rabi frequency.

Purely electric field manipulation of a HH spin-qubit via electric dipole spin resonance can also produce MHz frequencies thanks to the presence of the acceptor T$_d$ symmetry and spin-orbit Rashba terms. The required lack of inversion symmetry can be obtained by applying a vertical electric field. The HH-LH mixing given by this electric field and the T$_d$ symmetry term is also very important. The Rabi frequencies are proportional to the asymmetry of the hole wave function. A large Bohr radius (or smaller binding energy) allows an easier manipulation of the wave-function by electric field. The HH-LH mixing is inversely proportional to the strain, hence large strains are not desirable as they reduce the Rabi frequency.

 The g-TMR method is the one that gives the best Rabi frequencies for HH qubits. In exchange, this method is also the most exposed one to charge noise. The quantum well composition is not very important as long as the g-factors are not suppressed (this happens around $x\approx 0.4$) and the barrier is not very high. The best Rabi frequencies are obtained when the acceptor is close to one barrier and pushed by an electric field to the opposite barrier. At the sweet spots, however, the Rabi frequencies are much smaller but still larger than those obtained through ESR or EDSR.

 The LH qubit, can be easily manipulated by electrical means with frequencies of the order of GHz around the sweet spots, allowing both fast manipulation and good coherence properties. 

In general, the presence of the spin-orbit terms due to the acceptor (T$_d$ symmetry) and lack of inversion symmetry (Rashba type) allow the possibility of several ways of manipulating the spin state of a hole bound to an acceptor inside a quantum well. Due to the HH-LH splitting in the HH case, the relaxation and coherence properties are expected to be optimal, making this type of quantum wells good candidates as quantum memories. The LH ground states have good properties of coherence and still high manipulatibility near the sweet spots, making this type of quantum wells good candidates for quantum computation. This type of devices would also have good compatibility with electrically defined quantum dots, allowing the possibility of hybrid dot-acceptor qubits.

\acknowledgements
We thank useful discussions with D. Culcer, J. Salfi, and X. Hu. We acknowledge funding from Ministerio de Econom\'ia, Industria y Competitividad (Spain) via Grants No FIS2012-33521 and FIS2015-64654-P. JCAU thanks the support from "Ayudas para contratos predoctorales para la formaci\'on de doctores 2013", grant BES-2013-065888.

\bibliography{qwacceptors}

\end{document}